# A Generic Multi-node State Monitoring Subsystem

James A. Hamilton
*SLAC, Stanford, CA 94025, USA*

Gregory P. Dubois-Felsmann
*California Institute of Technology, CA 91125, USA*

Rainer Bartoldus
*SLAC, Stanford, CA 94025, USA*

*(for the Babar Computing Group)*

The BaBar online data acquisition (DAQ) system includes approximately fifty Unix systems that collectively implement the level-three trigger. These systems all run the same code. Each of these systems has its own state, and this state is expected to change in response to changes in the overall DAQ system. A specialized subsystem has been developed to initiate processing on this collection of systems, and to monitor them both for error conditions and to ensure that they all follow the same state trajectory within a specifiable period of time. This subsystem receives start commands from the main DAQ run control system, and reports major coherent state changes, as well as error conditions, back to the run control system. This state monitoring subsystem has the novel feature that it does not know anything about the state machines that it is monitoring, and hence does not introduce any fundamentally new state machine into the overall system. This feature makes it trivially applicable to other multi-node subsystems. Indeed it has already found a second application beyond the level-three trigger, within the BaBar experiment.

## 1. INTRODUCTION

The BaBar experiment at SLAC ran for its first three years with a makeshift mechanism for starting and monitoring the multiple processes that comprise the level-three trigger. This mechanism involved running sixty remote shell commands, each of which created an xterm window to display the output of the corresponding process. There was no way for the run control system to actually know the state of these processes. This mechanism was not only inefficient, but also introduced undesirable race conditions into startup, and sometimes caused cleanup problems due to the lack of positive control in error conditions.

To remedy this situation, a new mechanism was developed to start and monitor the level-three trigger farm and to maintain a consistent state across the multiple instances. This architecture of this mechanism was made quite general, in that it does not depend upon the detailed states of the processes it monitors. As such it can be described as a generic multi-node state monitoring subsystem. As a result the same monitoring subsystem can be used in multiple similar applications involving different target processes with different state machines.

## 2. EXISTING ONLINE STRUCTURE

### 2.1. General Description

Figure 1 is a simplified diagram of the BaBar online data acquisition system (DAQ) as it existed prior to July 2002. The major subdivisions depicted include run control, dataflow, detector controls and level 3 trigger. Run control implements overall control and monitoring of the DAQ, and provides a user interface for this purpose. Detector controls provides inputs for monitoring the physical status of the detector subsystems. Dataflow reads the raw data from the detector subsystems, processes it through a variety of states, and delivers events to the level-three trigger for further processing.

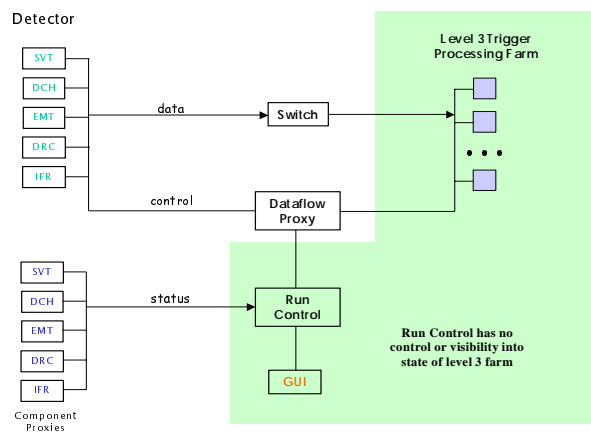

Figure 1: Pre-existing BaBar Online Structure

### 2.2. Run Control

The work described here falls most clearly within the scope of the run control system, so some further description of that system is necessary.

The run control system is based on the State Management Interface (SMI) [1] and the Distributed Information Management (DIM) system [2], both developed at CERN. SMI supports a collection of abstract state machines, expressed in the State Machine Language (SML). These abstract state machines communicate with, and operate on, the outside world via "proxies", which are also state machines, but whose implementation is usually expressed in C++ rather than SML.

DIM provides the communication subsystem by which the proxies communicate with the abstract state machines of run control. DIM is a relatively thin layer over TCP/IP, which offers a centralized name service and a robust disconnect and reconnect capability for failure recovery.

## 3. DEFICIENCIES IN THIS SYSTEM

Although the existing system did allow the BaBar experiment to take data for three years, it did have some deficiencies, as indicated in figure 1. Specifically, the states of the level-three trigger nodes are unknown to run control. In addition, it is not possible for run control to reset or restart level-three nodes when various error conditions arise.

For example, the dataflow system cannot proceed to the CONFIGURED state until level-three has reached the ALLOCATED state (the actual details of these states are not germane to this paper). But since run control cannot see the ALLOCATED state, human operator intervention was required in order to allow the system to proceed.

Similarly, it was sometimes necessary to manually login to individual level-three farm machines in order to clean up after certain kinds of system failures.

### 3.1. Alternatives for Correction

Here we describe several alternate ways to correct these deficiencies.

- Use the dataflow proxy. As can be seen from the figure, the dataflow proxy already communicates with the level-three nodes, so it is capable of performing the monitoring and control functions itself. Although this seems like an attractive solution, it turns out that by the time this work was undertaken, dataflow was already implemented. Because it is a rather complex subsystem already, it would have been potentially destabilizing to re-architect it to incorporate these additional functions.
- A second alternative was to implement each node as a separate run control proxy, and express the overall aggregation of the states using abstract SMI objects. This approach was feasible, but we wanted certain features that were awkward to achieve within SMI. In particular we wanted to allow dynamic variability in the number of nodes that actually participated in a given operation. In addition, we wanted timeouts and conflict detection to apply to each aggregate state transition.
- These considerations prompted us to take a third approach. This involved the implementation of a single new run control proxy, whose job is to monitor and control the states of the level-three farm and report a single resultant state back to run control.

## 4. THE MULTI-NODE PROXY

Figure 2 shows the BaBar online structure again with the new state monitoring proxy added. Figure 3 shows just the detail of the state monitoring subsystem itself.

Here we see the introduction of a manager dæmon process on each farm node that communicates with the central proxy via the DIM subsystem, and with the process it manages via a Unix pipe.

In the following description you will see that the architecture and implementation of the multi-node proxy are independent of the specifics of both BaBar run control and of the level-three trigger processing that it monitors.

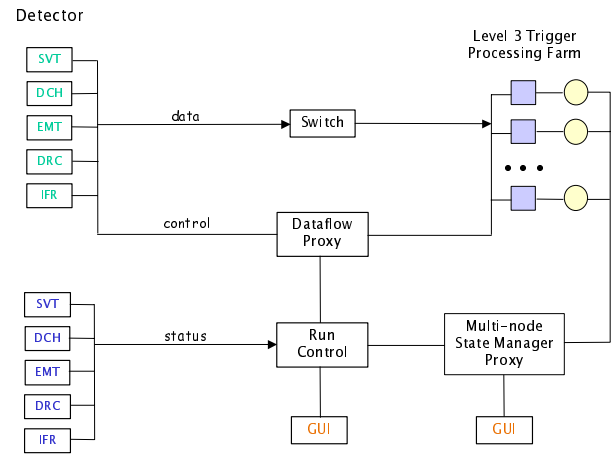

Figure 2: BaBar online structure with multi-node state manager

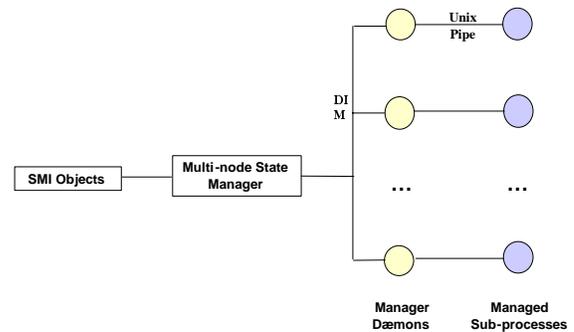

Figure 3: Detail of the Multi-node State Manager

### 4.1. Manager Operation

The goals of the central manager are as follows:

- Accept commands from the SMI state machine (e.g. run control) and pass them on to the dæmons, which in turn use them to control the managed process.
- Monitor the state transitions of the dæmons and report back a single aggregate state to the SMI state machine.
- Ensure that the state transitions are consistent among the multiple dæmons. If one dæmon goes to

state X, then all of the dæmons must go to state X, and must not visit any other states along the way.

- Ensure that dæmon state transitions are timely, both in response to commands, and in response to state changes initiated by other dæmons.
- Report error conditions back to the SMI state manager. This includes both errors detected by the dæmons and reported to the manager, but also dæmon misbehavior detected by the manager, as described above.

Although the manager must report a state to the SMI state manager, it does not itself constitute a new state machine within the system. Instead, with two exceptions, it simply adopts the states of the dæmons. This allows the same manager to be used with alternative sets of dæmons having differing state machines. And as we will see shortly, the state machines of the dæmons are also easily configurable. This is the primary reason for applying the term "generalized" to the multi-node state manager.

The two exceptions just mentioned are:

- The manager understands the initial, or inactive, state of the subsystem, which is called READY. All of the dæmons must have this as their initial state as well.
- The manager defines the state ERROR. It enters this state whenever it detects misbehavior or disappearance of the dæmons. It also enters this state whenever the dæmons report error states of their own.

All commands from run control are passed on to the dæmons. However, the manager also adds its own special semantics to two commands:

- START – this command establishes that there are sufficient dæmons available to activate the system, selects a subset of the available dæmons, and marks them as active. They remain active until they report the READY state. Subsequent commands are passed only to active nodes, and only active nodes are expected to participate in state transitions.
- RESET – this command tells the manager to expect READY as the next state reported by all dæmons. This is typically issued as part of error recovery, but may be issued as an abort operation at any time. The dæmons are expected to do whatever is necessary to return to their initial state, including terminating their sub-processes, and so on.

## 4.2. Dæmon Operation

As mentioned above, the dæmons are each controlled by a state machine, and it is these states that are reported to the manager and that ultimately become the states of the manager itself.

The state machine of the dæmons is straightforward and fairly traditional. It is based on a subsystem called the State Machine Framework, described in [3]. For our purposes we simply note that it provides for a list of states, transitions, and actions, and that it specified in a text file that is processed by the dæmon executable during initialization. The dæmon itself provides the list of available transitions and actions, which we enumerate here.

The transitions defined by the dæmon include:

- Commands from the manager, such as START and RESET, and also including network disconnect.
- Termination status from the sub-process, distinguished by exit code.
- A predefined set of events that occur within the sub-process and are reported through the Unix pipe to the dæmon.

The actions available to the state machine include starting a process, killing a process, cleanup after errors and shutting down the dæmon. There is currently no provision for passing actions to the sub-process.

## 4.3. State Reporting

When a dæmon changes state, it generally reports this state change to the manager. These reports also include a color name that can be used to display the new state in the manager's GUI.

The dæmon classifies its states into one of the following four types:

- Major: states that ultimately become states of the manager and get reported to run control. This manager state change will occur once all of the active dæmons have reported this state. The READY state is a major state by definition. The other major state are typically normal operational states, such as ALLOCATED, CONFIGURED and RUNNING, which originate in the sub-process that the dæmon is monitoring.
- Minor: states that may be of interest to the operator but do not become states of the manager itself, and are not reported to run control. These states are simply displayed in the manager's GUI. Examples of such states in the BaBar system are CONNECTING, and MAPPED.
- Micro: transient states that are of interest only to the internal operation of the dæmon, and are not reported at all to the manager. They are not expected to persist for more than a fraction of a second.
- Error: unexpected states, such as a non-zero exit code from the monitored process, are reported to the manager as distinct states for display in the GUI. However, the manager treats all error states as equivalent, and may respond by entering its own single error state.

## 4.4. Error Processing

Error processing is one of the more complex and interesting features of the multi-node manager. The most common type of error condition is the report of an error state from a dæmon. The manager is typically configured to allow a certain number of these error reports before declaring that the manager itself is in the ERROR state.

The dataflow subsystem is capable of "trimming" failed level-three trigger processes, and continuing to operate.

In addition to error reports from the dæmons, the manager detects many other kinds of error condition, including:

- Network disconnect of a dæmon.  If a dæmon node, or the dæmon itself crashes, or the network connection goes down for some reason, the dæmon will be marked as dead, inactive, and unavailable.  Because it is inactive, it is not expected to participate in future state changes.   In other respects, however, this is treated as equivalent to an error state reported by the dæmon.

- Transition to a conflicting state.  This condition is detected whenever a subset of the active dæmons reports state "A", followed by a dæmon reporting state "B".  This detection is complicated by the fact that dæmon state change sequences can occur very rapidly, and some dæmons may report state B after reporting state A, but before all other dæmons have reported state A.  This must be distinguished from the case where state B is reported *without* first reporting A.  This latter case causes the manager to enter ERROR, but the early reports are simply held pending until the in-progress transition completes.

- Timeout in state transition.  When the first report of a dæmon state change arrives while the manager is in a known state (i.e. all active dæmons are in the same state) a timer is started.  If this timer expires before all of the dæmons have reached the same new state, the manager enters the ERROR state.

- Timeout on START and RESET commands.  Similarly, when the main state manager issues one of these commands, a timer is started.  All dæmons must report some consistent state change before this timer expires.

When the manager enters the ERROR state no further action is taken until a RESET command is sent from the main state manager (e.g. run control) to clear the error.  During this time, dæmons may continue to operate and to report state changes, which are displayed but otherwise ignored, except that a state change to READY always marks a node as inactive.

Another kind of dæmon misbehavior is that it becomes unresponsive, even though it is still connected.   This condition is detected whenever a node fails, after a timeout, to return to READY following a RESET command.  Such nodes are marked unavailable and will not participate in future operations.

A final kind of misbehavior is detected when a dæmon reports a state transition while it is inactive.  When this happens, the node is marked unavailable, and the manager will issue a RESET command to attempt to get the node back to the READY state.  No overall manager error results from this action.

## 4.5.  Configurable Parameters

The manager has several configurable parameters that can be set through its user interface.  These are:

- Minimum node count – the START command in general will require a minimum number of dæmon nodes to be available in order for the start to succeed.   If this number is not available, the manager will not issue the START, and will go to ERROR.

- Maximum node count – it may be desirable in some cases to activate only a subset of the available nodes in the START command.  The unused nodes are simply left inactive and will not participate in the operation.

- Maximum error count – when a dæmon reports an error, or disappears, the manager may allow operation to continue without entering its own ERROR state.  The maximum error count parameter provides an upper bound on how many error reports will be tolerated before the manager enters ERROR.

- Timeout – the manager times out both commands and asynchronous state transitions.  The length of the timeout is configurable.

## 4.6.  The Communication Subsystem

The manager and dæmons use DIM as their communication subsystem.  What was needed was a low-level network messaging system.  The direct use TCP/IP was considered.  However, DIM is already used within BaBar in support of the SMI-based run control system.  DIM is a fairly thin layer over TCP, and has some additional advantages.

First, it provides a central naming service, so that the manager and the dæmons can establish communications without the need for well-known host names, IP addresses, or TCP/IP port numbers.   Also no network broadcast operations are required.

Second, DIM provides a robust disconnect/reconnect mechanism that helps in the face of both network and host failures, as well as component restarts.

The DIM API was not ideal for this application.  What was desired was something more akin to a reliable datagram service, or simply the read/write stream semantics provided by TCP.   However, it was straightforward to adapt it to our needs, and this slight mismatch was more than offset by the advantages mentioned above.

The performance of DIM in this application proved to be more than adequate.

## 5.  THE MANAGER GUI

The multi-node state manager has its own graphical user interface, which is implemented in TCL/TK.   It communicates with the manager via a Unix pipe.

The main purpose of the GUI is to display the current states of all the nodes under management.  Each state has a distinctive color for this purpose.  The colors, like the states themselves, are defined by the dæmon state machine, so the manager has no built-in knowledge of the colors.

Each dæmon node is displayed as a button. The button displays the color of the state, as well as the state name and node name. The actions defined on the buttons allow the operation to kill or restart individual dæmons, as well as view their output logfiles.

The GUI also displays the overall manager state, and a one-line description of the most recent action leading to the current state.

The main menu of the GUI allows for setting the configurable parameters of the manager, which are described above.

Figure 4 shows an example of the user interface.

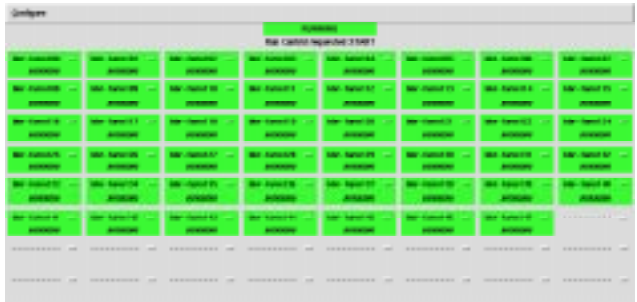

Figure 4: Example of the user interface in operation

## 6.  A SECOND APPLICATION

As it happens, an opportunity to test the generality of the multi-node state manager arose very quickly. This second application is shown in figure 5. This figure shows a portion of the BaBar fast monitoring subsystem. The purpose of this subsystem is to collect histograms from a random sampling of all level-one events in the detector, in order to monitor data quality in real time.

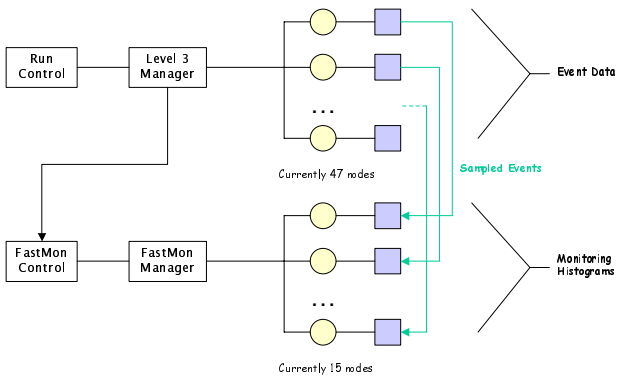

Figure 5: The Fast Monitoring Application

In order to monitor as much data as possible, multiple nodes are used to do the processing. Each of these nodes connects to one of the level-three nodes (managed by the main level-three manager) to obtain its sampling of level one events. We wrote a small SML program to control the overall operation of the fast monitoring subsystem, and this uses the multi-node state manager, unchanged, to control and monitor the individual histogram collection nodes. The controlling state machine of the fast monitoring subsystem is also able to monitor the state of the level-three manager to assist in sequencing its own operations.

## 7.  CONCLUSIONS

The goals of this work were to develop a subsystem to manage and monitor the states of multiple identical networked processes. It turned out that this could be done in a very general way that allowed the same manager to be used to control different subsystems with different state spaces. The result is a generalized multi-node state monitoring subsystem.

The monitoring subsystem has improved both the performance and the robustness of the BaBar online system. The performance was improved mainly by reducing the time to start a new set of processes, which had previously required remote shell invocation.

The robustness was improved in two ways. First, certain race conditions were eliminated because run control could now determine the overall state of the level-three processing farm. Second, when error conditions arise, cleanup is now much more reliable and predictable.

The new state manager has been in operation since November 2002, at the start of a new run. A second application has already been put in place, and has now been operational for several months.